\documentclass[12pt]{iopart}
\usepackage{iopams} 
\usepackage{graphics}
\usepackage{pennames}
\usepackage{amssymb}
\usepackage{setstack}
\begin{document}
\title[New results from the NA57 experiment]
{New results from the NA57 experiment}
\author{G E Bruno for  
the NA57 Collaboration~\footnote[1]{For the full author list see 
Appendix ``Collaborations'' in this volume.}
}
\address{
Dipartimento~IA~di~Fisica~dell'Universit{\`a}~e~del~Politecnico~di~Bari~and~INFN,~Bari,~Italy\\
}
\begin{abstract}
The production of hyperons in Pb-Pb and p-Be interaction at 40 $A$\ GeV/$c$\ 
beam momentum has been measured by the NA57 experiment. Strange particle 
enhancements at 40 $A$\ GeV/$c$\ are presented for the first time and 
compared to those measured at 158 $A$\ GeV/$c$.  
The transverse mass spectra of 
high statistics,
high purity samples of \PKzS, \PgL, $\Xi$\ and  $\Omega$\ particles 
produced  in Pb-Pb collisions at 158 $A$\ GeV/$c$\  
have been studied 
in the framework of the blast-wave model.    
The dependence of the freeze-out parameters on   
particle species  
and event centrality is discussed.  
\end{abstract}
\vspace{-0.6cm}
%
%
%
\section{Introduction} 
The experimental programme with 
heavy-ion beams at CERN SPS aims at the study of hadronic matter 
under extreme conditions of temperature, pressure and energy density.  

Within this programme, the WA97 experiment has measured  
an enhanced production of particles carrying one, two and three units  
of strangeness in central Pb-Pb collisions at 158 $A$\ GeV/$c$\  
with respect to proton  induced reactions at the same energy  
(Strangeness Enhancement)~\cite{WA97Enh}.   
The enhancement 
is defined as the hyperon yield per wounded nucleon in Pb-Pb collisions 
relative to the yield per wounded nucleon in p-Be collisions,  
where the number of {\em wounded nucleons} ($N_{wound}$) is the number of 
nucleons which are estimated to undergo at least   
one primary inelastic collision with another nucleon~\cite{WoundNucl}.  
The enhancement 
increases with the strangeness content 
of the hyperon: $\Omega$s are more enhanced than $\Xi$s which in turn  
are more enhanced than $\Lambda$s~\cite{WA97Enh}.  
This pattern was predicted more than 20 years ago as a consequence of  
Quark Gluon Plasma (QGP) formation~\cite{Rafelski}. The WA97 
results are not reproduced by any  
conventional hadronic microscopic model.  

The assessment of the combined results from a series of heavy-ion 
experiments suggests indeed that a new state of matter, which features many 
of the characteristics of the theoretically predicted QGP, 
is produced in central Pb-Pb collisions at 158 $A$\ 
GeV/$c$~\cite{CERN_ANN}.  

NA57 at the CERN SPS is a dedicated 
second-generation  
experiment for the study of the production of strange and multi-strange particles  
in Pb-Pb and p-Be collisions~\cite{NA57proposal}.   
It continues and extends the study initiated by its predecessor WA97, 
by {\em (i)} enlarging the triggered fraction of the inelastic  
cross-section, thus extending the centrality range towards less central  
collisions and {\em (ii)} collecting data also at lower (40 $A$\ GeV/$c$)  
beam momentum in order to study the energy dependence of the  
enhancements.  

In this paper we 
present for the first time results on strangeness enhancements 
at 40 $A$\ GeV/$c$. The results are then  
compared with those obtained at 158 $A$\ GeV/$c$.  

A detailed study of the  
transverse mass ($m_{\tt T}=\sqrt{p_{\tt T}^2+m^2}$) 
spectra   
for  \PgL, \PgXm, \PgOm\ hyperons, their antiparticles and \PKzS\ 
measured in Pb-Pb collisions at 158 $A$\ GeV/$c$, has been performed.  
The shapes of the $m_{\tt T}$\ spectra are expected to 
depend both on the thermal motion of the  
particles and on the collective flow driven by the pressure.  
To disentangle the two contributions 
we rely on 
the {\it blast-wave} model~\cite{BlastRef,BlastRef2},  
which assumes cylindrical symmetry for an expanding fireball in local  
thermal equilibrium, testing different hypotheses on the transverse 
flow profile. 
\section{Data sample and analysis}
The results presented in this paper are based on the analysis of the full data  
sample collected in Pb-Pb collisions, consisting of 460 M events  
at 158 $A$\ GeV/$c$\  
and 240 M  events at 40 $A$\ GeV/$c$. The analysed sample for p-Be collisions at  
40 $A$\ GeV/$c$\ consists of 110 M events.   
A separate, smaller sample (60 M events)  
has also been collected,  
its analysis is currently on the way to completion.   
At 158 GeV/$c$, we use as reference data those collected in p-Be and p-Pb  
interactions by WA97.  

\begin{figure}[b]
\centering
\resizebox{0.60\textwidth}{!}{%
\includegraphics{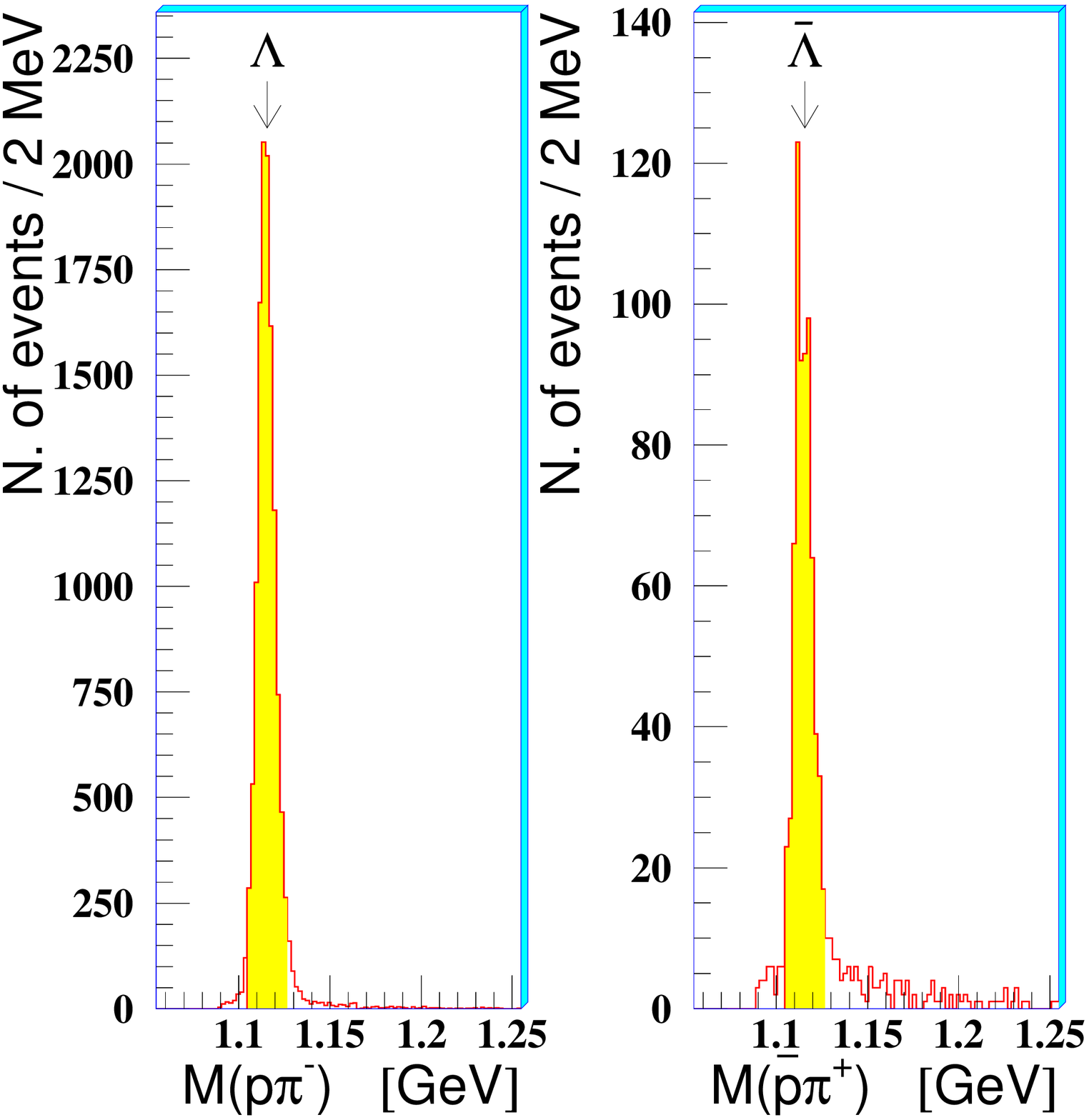}
\includegraphics{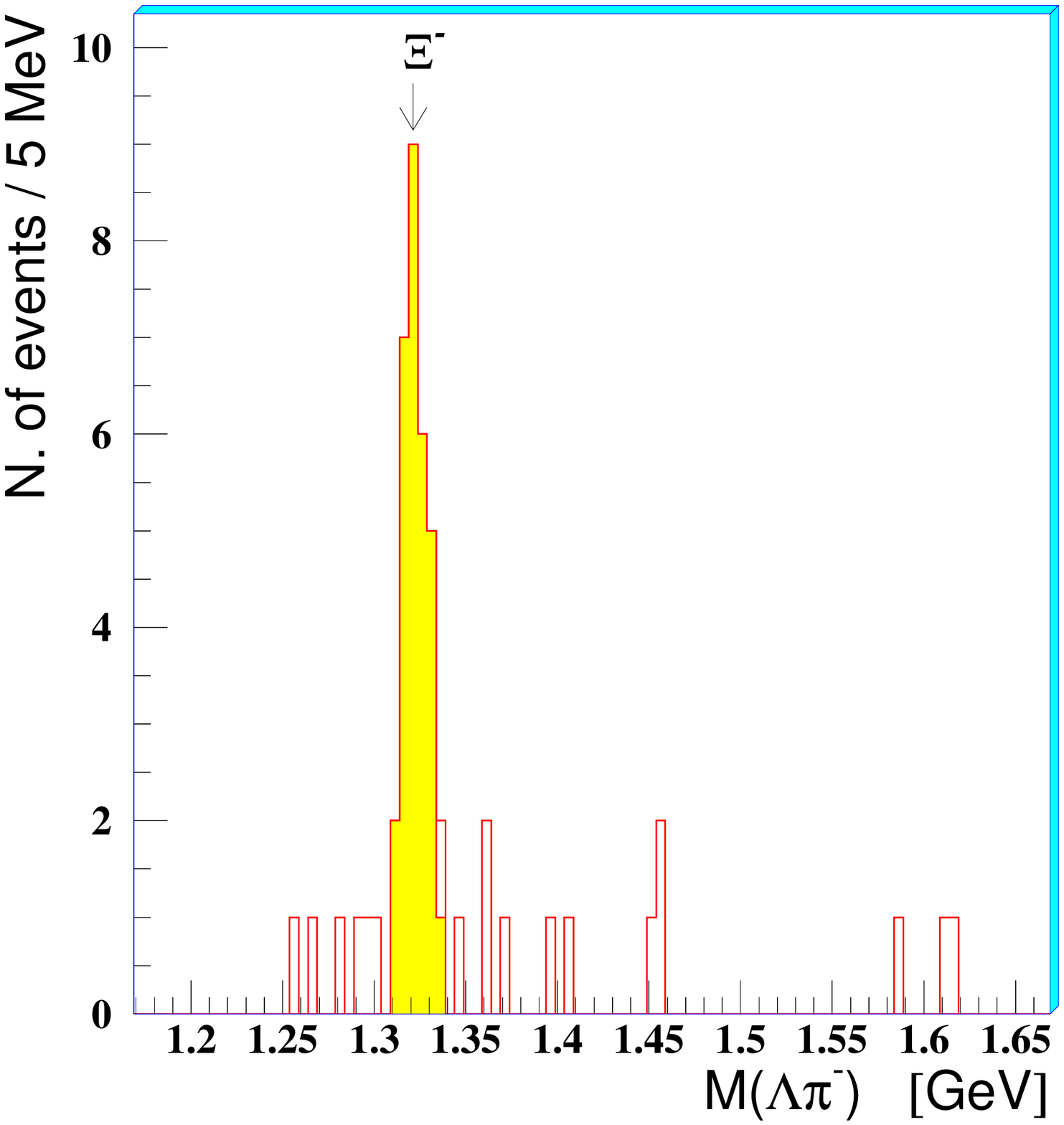}}\\
\caption{\rm Sample invariant mass spectra for 
	     \Pp\Pgpm, \Pap\Pgpp and \PgL\Pgpm\ in p-Be at 40 GeV/$c$. 
	}
\label{fig:signals}
\end{figure}

The NA57 apparatus  has been described in detail elsewhere~\cite{MANZ}.
The strange particle signals are extracted 
by reconstructing the weak   
decays into final states containing only charged particles,   
using geometric and kinematic constraints, 
with a method similar to that used in the WA97 experiment~\cite{WA97PhysLettB433}. 
The invariant mass spectra 
in p-Be at 40 $A$\ GeV/$c$\ 
for \Pp\Pgpm, \Pap\Pgpp\ and \PgL\Pgpm\  
after all analysis cuts  are shown in figure~\ref{fig:signals}.  
Hyperon  peaks are well centered at the PDG values~\cite{PDG}  
with FWHM of about 10 MeV (\PgL) and 15 MeV (\PgXm). 

For each particle species we define the fiducial  
acceptance window using a Monte Carlo simulation of the apparatus, 
in order to exclude the borders where the systematic errors  
are more difficult to evaluate.  
For all hyperons, the acceptance window for particle and  
antiparticle are the same,   
due to the reflection symmetry of our apparatus with 
respect to the magnetic field direction. 
As a further check the orientation of the magnetic field  
was periodically inverted.   

All data are corrected for geometrical acceptance and for detector and 
reconstruction inefficiencies on an 
event-by-event 
basis, with the 
procedure described in reference~\cite{QM02Manzari}. 
It has been checked that the experimental smearings in $p_{\tt T}$\ and $y$\  
have negligible effects on the weights.  
The simulation used for calculating the correction factors has been checked 
in detail (see, e.g., reference~\cite{Kristin}) 
by comparing real and Monte Carlo distributions  
for the selection parameters.  

As a measure of the collision centrality we use the number of wounded nucleons
$N_{wound}$\ computed via the Glauber model.   
The procedure for the measurement of the multiplicity distribution and  
the determination of the collision centrality for each class 
is described in reference~\cite{Multiplicity}.  
The distribution of the charged particle multiplicity  
measured in Pb-Pb interactions  
has been divided into five centrality classes (0,1,2,3,4), class 0 
being the most peripheral and class 4 being the most central.  
The average $N_{wound}$\ in the five centrality classes is   
given in table~\ref{tab:participants}.  
\begin{table}[h]
\caption{Average number of $N_{wound}$\ in  the five  classes defined  
in Pb-Pb interactions. 
\label{tab:participants}}
\begin{center}
\begin{tabular}{|c|c|c|c|c|}
\hline
                       $0$   &   $1$     &    $2$    &   $3$     &   $4$ \\ \hline
$58\pm4$ & $117\pm4$ & $204\pm3$ & $287\pm2$ & $349\pm1$ \\
\hline
\end{tabular}
\end{center}
\end{table}
\noindent
\vspace{-1.0cm}
\section{Strangeness enhancement}
The double-differential $(y,m_{\tt T})$\ distribution for each particle 
species has been parametrized using the expression  
\begin{equation}
\label{eq:expo}
\frac{d^2N}{m_{\tt T}\,dm_{\tt T} dy}=f(y) \hspace{1mm} \exp\left(-\frac{m_{\tt T}}{T_{app}}\right)
\end{equation}
assuming the rapidity distribution to be flat within our acceptance region
($f(y)={\rm const}$).
By using equation~\ref{eq:expo} we can extrapolate   
the yield measured in the selected acceptance window to a common phase space 
window covering full $p_{\tt T}$\ and one unit of rapidity centered 
at midrapidity:  
\begin{equation}
Y=\int_{m}^{\infty} {\rm d}m_{\tt T} \int_{y_{cm}-0.5}^{y_{cm}+0.5} {\rm d}y
  \frac{{\rm d}^2N}{{\rm d}m_{\tt T} {\rm d}y}.
\label{eq:yield}
\end{equation}
\noindent
The dependence of the \PKzS, \PgL, $\Xi$\  and $\Omega$\ yields 
on centrality and energy in Pb-Pb  collisions has been discussed  
in a separate contribution to the conference~\cite{DElia}.   
In p-Be collisions at 40 GeV/$c$\ we have measured  
the yields of \PgL, \PagL\ and \PgXm.  
For the \PagXp\ particle, due to the low cross-section and limited  
statisitcs, we 
could estimate only 
an upper limit to  
the production yield.  This enables us to put an upper limit to 
the \PagXp\ enhancement.  

\begin{figure}[tb]
\centering
\resizebox{0.70\textwidth}{!}{%
\includegraphics{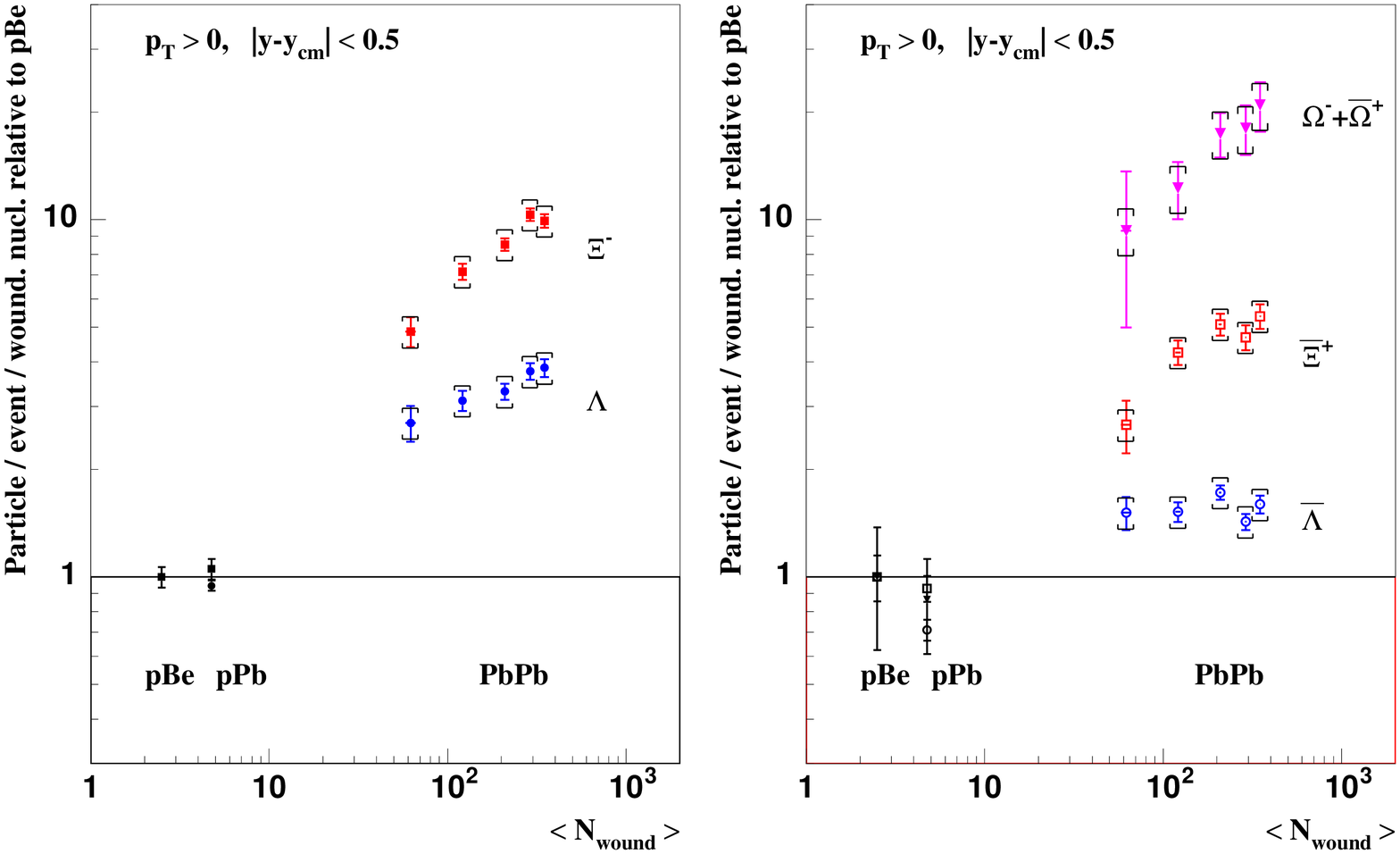}} \\
\resizebox{0.70\textwidth}{!}{%
\includegraphics{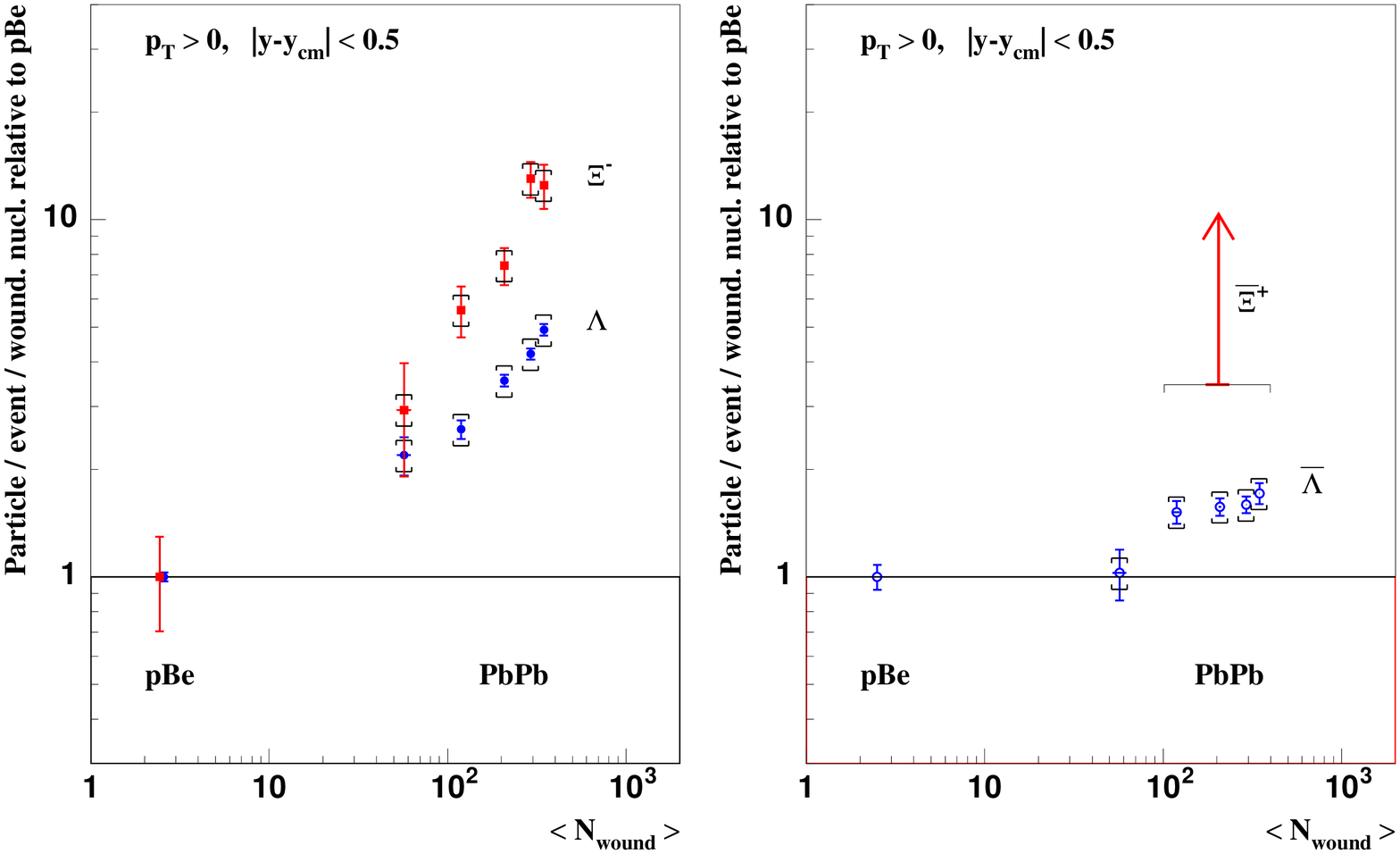}}
\caption{\rm 
              Hyperon enhancements $E$\  
	     as a function of the number of wounded
             nucleons at 158 (top) and at 40 (bottom) $A$\ GeV/$c$. The symbol
             $_{\sqcup}^{\sqcap}$\  shows the systematic error.}
\label{fig:HypEnh} 
\end{figure}

The 
enhancement $E$\ 
is defined as  
\begin{equation}
E={\left(  \frac{Y}{<N_{wound}>}  \right)_{Pb-Pb}} / {
   \left(  \frac{Y}{<N_{wound}>}  \right)_{p-Be}     }
\label{eq:enh2} 
\end{equation} 
In figure~\ref{fig:HypEnh} we show the 
enhancements 
as a function of $N_{wound}$\  
at 158 (top) and 
at 40 $A$\ GeV/$c$ (bottom).   
The enhancements are 
shown separately for particles containing at least 
one valence quark in common with the nucleon (left) and for those with 
no valence quark in common with the nucleon (right).  

The 158 $A$\ GeV/$c$\ results confirm the picture which emerged from 
WA97 --- the 
enhancement 
increases with the strangeness content of the hyperon --- 
and extend the measurements to lower centrality.   
In the Pb-Pb sample one sees a    
significant centrality dependence of the yields per wounded nucleon for all  
hyperons except for \PagL. However, for the two most central classes 
data are compatible with a saturation of the enhancements.   

A significant enhancement of strangeness production  
when going from p-Be to Pb-Pb  
is observed also in the 40 $A$\ GeV/$c$\ data. 
The arrow in figure~\ref{fig:HypEnh} indicates the lower limit to the \PagXp\   
enhancement 
in the four most central classes   
at 95\% confidence level.   
The enhancement pattern  
follows the same hierarchy with the strangeness content observed at 158 GeV/$c$:  
$ E(\PgL) < E(\PgXm)$\ and $E(\PagL) <  E(\PagXp)$.   

Comparing   
the measurements at the two  
beam momenta: for the most central collisions (bins $3$\ and $4$) the 
enhancements are 
higher at 40 than at 158 GeV/$c$,    
the increase with $N_{wound}$\ is steeper at 40 than at 158 GeV/$c$.  

The strangennes enhancement,  when  
described as a consequence of the transition from  the  
canonical to the asymptotic grand canonical limit, is indeed 
predicted to be a decreasing function of the collision energy~\cite{Redlich}.  
However, that model neither reproduces the steepness  
nor the amount for central collisions of 
the measured enhancements.  
\section{Transverse mass spectra in Pb-Pb at 158 $A$\ GeV/$c$}
\subsection{Exponential fits}
The inverse slope parameter $T_{app}$\ (``apparent temperature'')  
has been extracted by means of a maximum likelihood fit of equation~\ref{eq:expo} 
to the data. 
The  
apparent temperature is interpreted as due to the  
thermal motion coupled with a collective transverse flow  
of the fireball components~\cite{BlastRef,BlastRef2}.    

The $1/m_{\tt T} \, dN/dm_{\tt T} $\ distributions 
are well described by exponential functions~\cite{BlastPaper}.   
In the next section, we exploit some small deviations from  
the exponential in an attempt to disentangle the transverse collective flow 
from the thermal motion.  

The inverse slope parameters $T_{app}$\ 
are given  in table~\ref{tab:InvMSD}  as a function of centrality.  
\begin{table}[h]
\caption{
         Inverse slopes (MeV) of the  
         $m_{\tt T}$\ distributions for the five Pb-Pb centrality 
	 classes ($0$,$4$), and 
	 for p-Be and p-Pb interactions~\cite{Slope-p}.  
	 Only statistical errors are shown. In Pb-Pb, systematic errors are 
	 estimated to be 10\% for all centralities.   
\label{tab:InvMSD}}
\begin{center}
\footnotesize{
\begin{tabular}{|c|c|c|cc|c|cc|} 
\hline
      &  p-Be   &   p-Pb    &    0      &    1      &    2      &    3    &    4   \\ \hline
\PKzS &$197\pm4$& $217\pm6$ & $239\pm15$ & $239\pm8$ & $233\pm7$ & $244\pm8$ & $234\pm9$ \\
\PgL  &$180\pm2$& $196\pm6$ & $237\pm19$ & $274\pm13$ & $282\pm12$ & $315\pm14$ & $305\pm15$ \\
\PagL &$157\pm2$& $183\pm11$ & $277\pm19$ & $264\pm11$ & $283\pm10$ & $313\pm14$ & $295\pm14$ \\
\PgXm &$202\pm13$&$235\pm14$ & $290\pm20$ & $290\pm11$ & $295\pm9 $ & $304\pm11$ & $299\pm12$ \\
\PagXp&$182\pm17$&$224\pm21$ & $232\pm29$ & $311\pm23$ & $294\pm18$ & $346\pm28$ & $356\pm31$ \\
\PgOm+
\PagOp&$169\pm40$& $334\pm99$ & \multicolumn{2}{c|}{$274\pm34$} & $274\pm28$ &
        \multicolumn{2}{|c|}{$268\pm23$} \\
\hline
\end{tabular}
}
\end{center}
\end{table}
An increase of 
$T_{app}$\ 
with  
centrality is observed in Pb-Pb  
for \PgL, \PagXp\  and possibly also for \PagL. 
Inverse slopes for p-Be and p-Pb collisions~\cite{Slope-p}  
are also given in table~\ref{tab:InvMSD}. In central  
and semi-central Pb-Pb collisions (i.e. classes 1 to 4) one observes 
a baryon-antibaryon symmetry in the shapes of the 
spectra.  
This symmetry is not observed in p-Be collisions.  
The similarity of baryon and antibaryon $m_{\tt T}$\ slopes observed in Pb-Pb  
suggests that strange baryons and antibaryons are produced by a  similar   
mechanism.   
\subsection{Blast-wave description of the spectra}
%
The blast-wave model~\cite{BlastRef} predicts 
a double differential cross-section    
of the form:  
\begin{equation}
\frac{d^2N_j}{m_{\tt T} dm_{\tt T} dy} 
    = \mathcal{A}_j  \int_0^{R_G}{ 
     m_{\tt T} K_1\left( \frac{m_{\tt T} \cosh \rho}{T} \right)
         I_0\left( \frac{p_{\tt T} \sinh \rho}{T} \right) r \, dr}
\label{eq:Blast}
\end{equation}
where $\rho(r)=\tanh^{-1} \beta_{\perp}(r)$\ is a transverse boost,   
$K_1$\ and $I_0$\ are  modified Bessel functions, $R_G$\ is the 
transverse geometric radius of the source at freeze-out 
and $\mathcal{A}_j$\ is a normalization constant.  
%
The transverse velocity field $\beta_{\perp}(r)$\ has been parametrized 
according to a power law: 
\begin{equation}
\beta_{\perp}(r) = \beta_S \left[ \frac{r}{R_G} \right]^{n}  
  \quad \quad \quad r \le R_G
\label{eq:profile}
\end{equation}  
With  
this type of profile the numerical value of $R_G$\ does not 
influence the shape of the spectra but just the absolute  normalization 
(i.e. the $\mathcal{A}_j$\ constant). 
%
The parameters which can be extracted from a fit of equation~\ref{eq:Blast} to 
the experimental spectra are thus the thermal freeze-out 
temperature $T$\ and the 
{\em surface} 
transverse flow velocity $\beta_S$. 
Assuming a uniform particle density, the 
latter can be replaced by the {\em average} transverse flow 
velocity,  
$
<\beta_{\perp}> = \frac{2}{2+n}  \beta_S
$.   

\begin{figure}[bt]
\centering
\resizebox{0.90\textwidth}{!}{%
\includegraphics{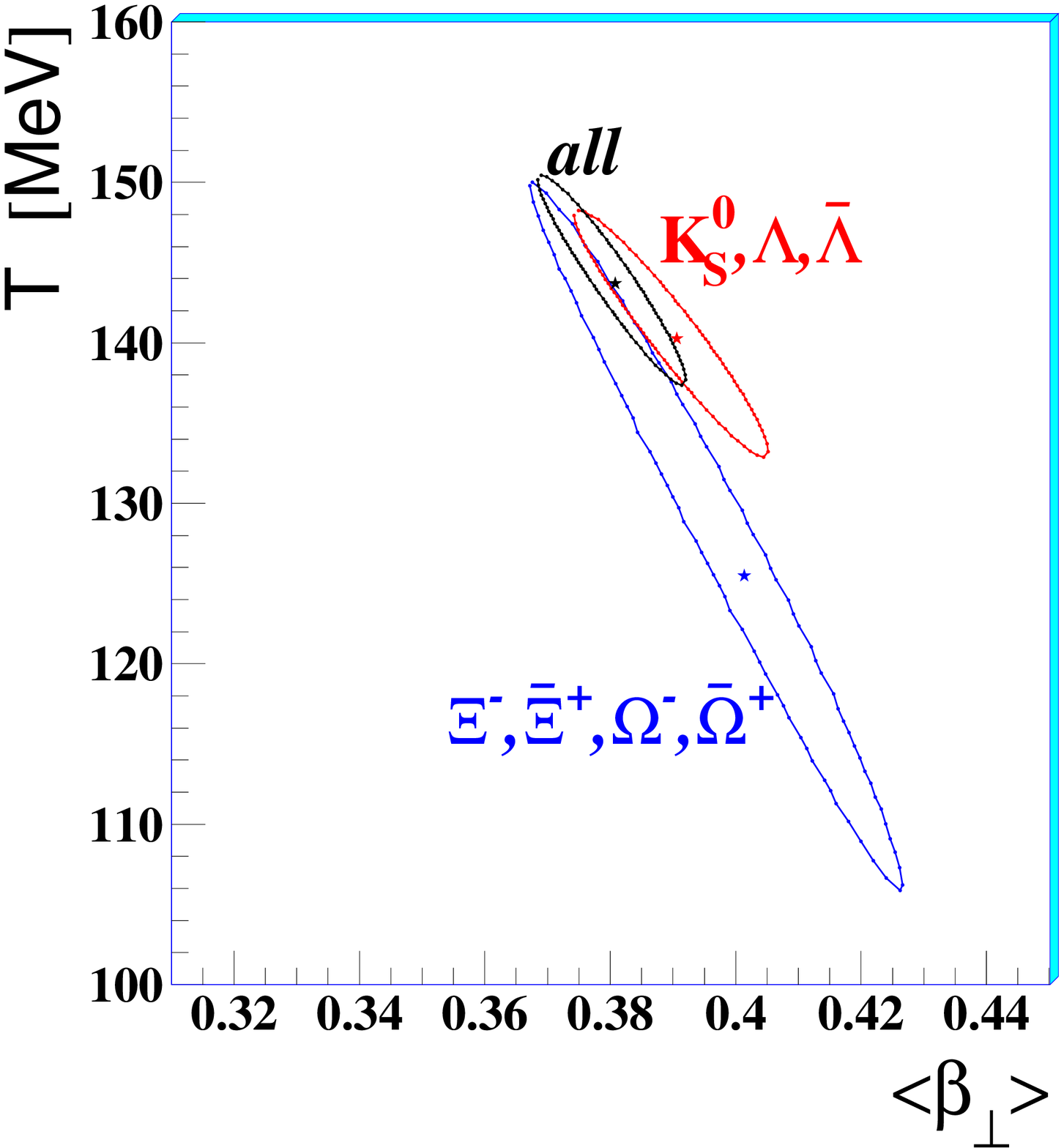}
\includegraphics{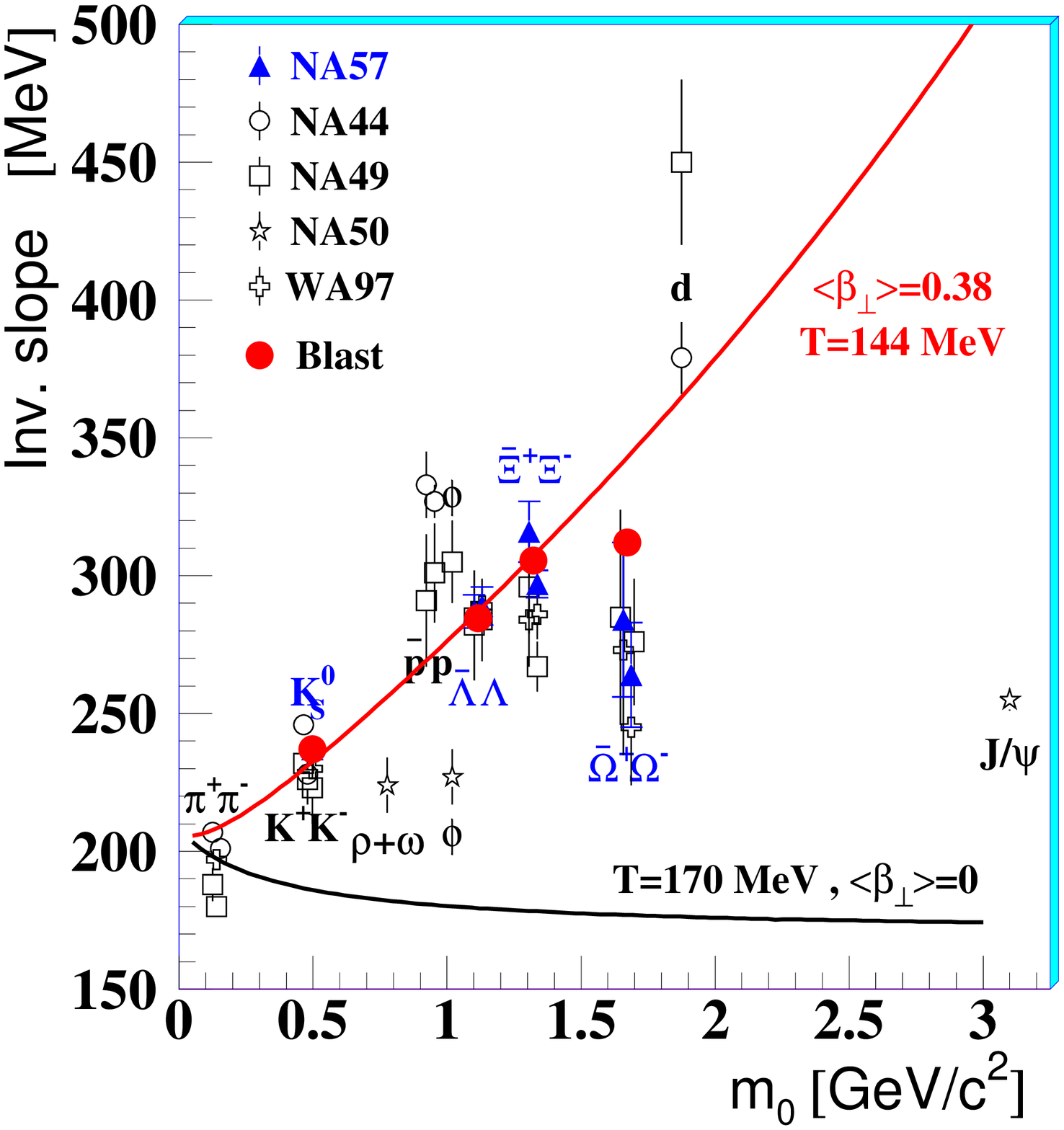}}
\caption{Left: the thermal freeze-out temperature vs the average transverse flow 
         velocity for blast-wave fits using a linear ($n=1$) velocity profile.   
         The 1$\sigma$\ contours are shown, with the markers indicating  
	 the optimal fit locations.  
	 Right: prediction of the blast-wave model for  
         inverse slopes (see text for details). For data references 
	 see~\cite{BlastPaper}.}  
\label{fig:BlastPred}
\end{figure}

The global fit of equation~\ref{eq:Blast} with $n=1$\ 
to the data points of all the measured strange particle spectra 
successfully describes all the distributions 
with $\chi^2/ndf=37.2/48$, yielding the following values for the two parameters  
$T$\ and $ <\beta_\perp>$:  
\[ \fl
 T = 144 \pm 7 {\tt (stat)} \pm 14 {\tt (syst)} {\rm MeV} \, , \quad 
 <\beta_\perp>=0.381 \pm 0.013 {\tt (stat)} \pm 0.012 {\tt (syst)} \; .
\nonumber 
\]
The $T$\ and $<\beta_\perp>$\ parameters are found to be statistically 
anti-correlated, as can be seen from the confidence level contours shown    
in figure~\ref{fig:BlastPred}.  The systematic errors on $T$\ 
and $<\beta_{\perp}>$\  are instead correlated;  
they are estimated to be $10\%$\ and $3\%$, respectively.  
The use of the three profiles $n=0$, $n=1/2$\ and $n=1$\ results in     
similar values of the freeze-out temperatures and of the  average transverse 
flow velocities, with good values of $\chi^2/ndf$. 
The quadratic profile is disfavoured by our data~\cite{BlastPaper}.  

It has been suggested (e.g.~\cite{Hecke}),   
based on WA97 results   
on the hyperon $m_{\tt T}$\ slopes~\cite{MtWA97}     
compatible with those of NA57, that the thermal  
freeze-out occurs earlier for $\Omega$\ and possibly for $\Xi$\ than for 
particles of strangeness 
$0$, due to the low scattering  cross-sections for $\Omega$ and $\Xi$.  
The 1$\sigma$\ contours of the separate blast-wave fits for singly  
and multiply strange particles are shown in figure~\ref{fig:BlastPred}. 
The results of the fits for both 
groups of particles are compatible with the result of the global fit 
determination.   
However, the fit for the multiply strange particles is statistically  
dominated by the $\Xi$;  in fact the $\Xi+\Omega$\  
contour remains essentially unchanged when fitting the $\Xi$\ alone.  
\newline
For the $\Omega$, due to the lower statistics,  
it is not possible to extract significant values  
for both freeze-out parameters  
from its spectrum alone  
(as can be done for the $\Xi$).   
Any possible deviation for the $\Omega$\   
from the freeze-out systematics extracted from the combined fit to the 
$\PKzS$, $\Lambda$\ and $\Xi$\  spectra  
can only be inferred from the integrated information of the 
$\Omega$\ spectrum, i.e. from its inverse slope.  
In figure~\ref{fig:BlastPred} we plot a compilation of inverse slopes 
measured in Pb-Pb collisions at 158 $A$\ GeV/$c$,  
superimposed to blast-wave model results.   
The full lines represent the inverse slope one would obtain by fitting 
an exponential to a ``blast--like'' $1/m_{\tt T} \, dN/dm_{\tt T}$\ 
distribution for a generic particle of mass $ m_{0} $,    
in the common range $ 0.05 < m_{\tt T} - m_{0}  < 1.50$\ GeV/$c^2$,   
for two different freeze-out conditions:   
absence of transverse flow ($<\beta_{\perp}>=0$) and our best fit determination.  
Since the inverse slope is also a function of the $m_{\tt T} - m_{0}$\ 
range where the fit is performed, we have also computed the blast-wave 
inverse slopes of $\PKzS$, $\PgL$, $\Xi$\ and $\Omega$\  
spectra in the $m_{\tt T} - m_{0}$\ ranges of NA57  
(closed circles).  
The measured values of the inverse slope of the $\Omega$\  deviate    
from the blast-wave trend of the other strange particles.  

\begin{figure}[t]
\centering
\resizebox{0.40\textwidth}{!}{%
\includegraphics{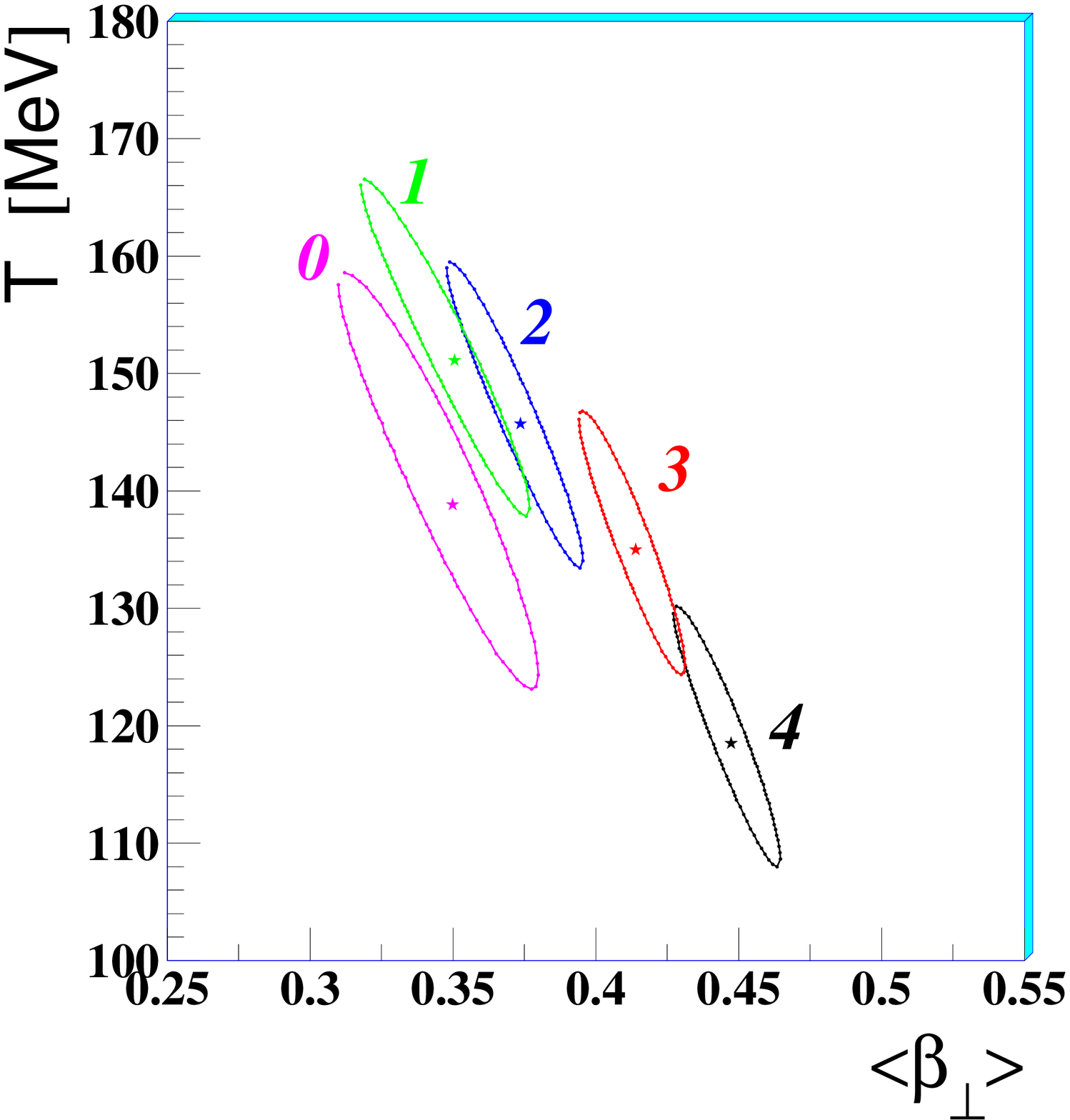}}
\caption{\rm The 1$\sigma$\ confidence level contours from
               fits in each centrality class.} 
\label{fig:cont_msd}
\end{figure}

We have also performed the global fit to the spectra  
for each of the five centrality classes defined in table~\ref{tab:participants}.  
In figure~\ref{fig:cont_msd} we show the $1\sigma$\ confidence level contours 
as obtained for the $n=1$\  profile.   
The observed trend is as follows:  
the more central the collisions the larger the transverse collective 
flow and the lower the final thermal freeze-out temperature.  
Higher freeze-out temperatures for more peripheral collisions may be interpreted 
as the result of an earlier decoupling of the expanding system.  
\section{Conclusions}
We have reported an enhanced production of \PgL, \PagL, \PgXm\ and \PagXp\ 
when going from p-Be to Pb-Pb collisions at 40 $A$\ GeV/$c$.  
The enhancement pattern follows the 
same hierarchy with the strangeness content as at 158 GeV/$c$: 
$ E(\PgL) < E(\PgXm)$, 
$ E(\PagL) < E(\PagXp)$. 
For central collisions (classes $3$\ and $4$) the enhancement is larger 
at 40 GeV/$c$.  
In Pb-Pb collisions the hyperon yields increase with $N_{wound}$\ faster at 40 than at 
158 $A$\ GeV/$c$.  

The analysis of the transverse mass spectra at 158 $A$\ GeV/$c$\ 
in the framework of the blast-wave  model 
suggests  that after a central 
collision the system expands explosively and then it 
freezes-out when the temperature is of the order  
of 120 MeV with an average transverse  flow   
velocity of about one half of the speed of light.  
The inverse slope of the $\Omega$\ particle deviates from the prediction of  
the blast-wave model tuned on other strange particles (\PKzS, \PgL\ and $\Xi$).    

Finally, the  results on the centrality dependence of the  
expansion dynamics  
indicate that with increasing centrality  
the transverse flow velocity increases and the final temperature decreases.  
%
%


\section*{References}
\begin{thebibliography}{33}
%
%
\bibitem{WA97Enh}  Andersen E {\it et al.} 1999 \PL B {\bf 449} 401  \nonum
                   Antinori F {\it et al.} 1999 \NP A {\bf 661} 130c
\bibitem{WoundNucl} Bialas A, Bleazy\'nski M and Czy\.{z} W 1976 \NP B {\bf 111} 461   
\bibitem{Rafelski} Rafelski J and M\"{u}ller B 1982 \PRL {\bf 48} 1066  \nonum
                   Rafelski J and M\"{u}ller B 1986 \PRL {\bf 56} 2334
\bibitem{CERN_ANN} Heinz~U and Jacob~M 2000 {\it Preprint} nucl-th/0002042 and 
                   reference therein 
\bibitem{NA57proposal} Caliandro R {\it et al.}, NA57 proposal, 1996 
{\it CERN/SPSLC 96-40, SPSLC/P300} 
\bibitem{BlastRef} Schnedermann E, Sollfrank J and Heinz U 1993 \PR C
                   {\bf 48} 2462
\bibitem{BlastRef2} Schnedermann E, Sollfrank J and Heinz U
                   1994 \PR C {\bf 50} 1675
%
%
\bibitem{MANZ} Manzari V {\it et al.} 1999 \JPG {\bf 25} 473   \nonum
               Manzari V {\it et al.} 1999 \NP A {\bf 661} 761c 
%
%
\bibitem{WA97PhysLettB433} Andersen E {\it et al.} 1998 \PL B {\bf 433} 209 \nonum
                           Lietava R {\it et al.} 1999 \JPG {\bf 25} 181 \nonum
                           Fini R A {\it et al.} 2001 \JPG {\bf 27} 375
\bibitem{PDG} Hagiwara K {\it et al.} 2002 \PR D {\bf 66} 010001
\bibitem{QM02Manzari} Manzari V {\it et al.} 2003 \NP A {\bf 715} 140c
\bibitem{Kristin} Fanebust K {\it et al.} 2002 \JPG {\bf 28} 160
\bibitem{Multiplicity} Carrer N {\it et al.} 2001 \JPG {\bf 27} 391
\bibitem{DElia} Elia D {\it et al.} 2004 \JPG,  theese proceedings
\bibitem{Redlich} Tounsi~A, Mischke~A and Redlich~K 2003 \NP A {\bf 715} 565c
%
%
\bibitem{BlastPaper} Antinori F {\it et al.} 2004 \JPG, in press    
                     ({\it Preprint } nucl-ex/0403016)  
\bibitem{Slope-p} Fini R A {\it et al.} 2001 \NP A {\bf 681} 141c
%
%
\bibitem{Hecke} van~Hecke~H, Sorge~H and Xu~N 1998 \PRL {\bf 81} 5764
\bibitem{MtWA97} Antinori F {\it et al.} 2000 {\it Eur. Phys. J.}
                C {\bf 14} 633
\end{thebibliography}
\end{document}